# RADIATION DAMAGE AND RECOVERY OF CRYSTALS: FRENKEL VS. SCHOTTKY DEFECT PRODUCTION

V. I. Dubinko**,**

NSC Kharkov Institute of Physics and Technology NAS of Ukraine,

Akademicheskaya Str.1, Kharkov 61108, Ukraine

**ABSTRACT**

A majority of radiation effects studies are connected with creation of radiation-induced defects in the crystal bulk, which causes the observed degradation of material properties. The main objective of this chapter is to describe the mechanisms of recovery from radiation damage, which operate during irradiation but are usually obscured by the concurrent process of defect creation. Accordingly, the conventional rate theory is modified with account of radiation-induced Schottky defect formation at extended defects, which often acts against the mechanisms based on the Frenkel pair production in the crystal bulk. The theory is applied for the description of technologically and fundamentally important phenomena such as irradiation creep, radiation-induced void "annealing" and the void ordering.

# Content

## 1   INTRODUCTION

Radiation damage in crystalline solids originates from the formation of Frenkel pairs of vacancies and self-interstitial atoms (SIAs) and their clusters. The difference in the ability to absorb vacancies and SIAs by extended defects is thought to be the main driving force of microstructural evolution under irradiation. A recovery from radiation damage is usually observed under sufficiently high temperatures and it is driven by thermal fluctuations resulting in the evaporation of vacancies (Schottky defects) from voids and dislocations and in the fluctuation-driven overcoming of obstacles by gliding dislocations. These recovery mechanisms can be efficient only at sufficiently high temperatures. However, there is increasing evidence that the so-called "thermally activated" reactions may be modified under irradiation. Thus, in refs. [1-3] it was demonstrated that Schottky defects could be produced not only by thermal fluctuations but also by interactions between extended defects and *excitons* in di-atomic ionic crystals. Excitons are the excitations in the electron sub-system, which can be formed under irradiation in *insulators* and move by diffusion mechanism towards extended defects thus providing a mechanism of Schottky defect formation at the defect surfaces. Well established excitations in ionic system include unstable Frenkel pairs (UFPs) [4, 5] and focusing collisions ("*focusons*") [6-9]. More recently, essentially non-linear lattice excitations, called *discreet breathers* (DBs) or *quodons* have been considered [10-14]. Accordingly, the rate theory of microstructure evolution in solids has been modified with account of the production of Schottky defects at voids and dislocations due to their interaction with the radiation-induced lattice excitations [15-21]. In this chapter, recent developments and applications of the new theory are presented.

The chapter is organized as follows. In the section 2, the competition between the Frenkel and Schottky defect production is illustrated at such technologically important phenomenon as irradiation creep.

In the sections 3 and 4, original experimental data on the radiation-induced void "annealing" are presented and analyzed in the framework of the rate theory modified with account of the radiation-induced Schottky defect formation.

In the section 5, we consider one of the most spectacular phenomena in physics of radiation effects, namely, the void lattice formation [22-28]. It is often accompanied by a saturation of the void swelling with increasing irradiation dose, which makes an understanding of the underlying mechanisms of both scientific significance and practical importance for nuclear engineering. We compare the popular mechanisms of void ordering based on anisotropic *interstitial transport* [22-27] with the original mechanism based on the anisotropic *energy transfer* provided by quodons [28].

## 2 IRRADIATION CREEP

Irradiation creep is one of the most outstanding puzzles of the theory of radiation damage, which has been studied rather extensively due to its technological importance [29-36]. It is known that under typical neutron fluxes, irradiation creep dominates over thermal creep below about 0.5 $T_m$ and shows rather *weak or no dependence on irradiation temperature*. However, for a long time, even this general trend was not well understood.

Conventional creep models are based on the so called *stress-induced preferential absorption* (SIPA) of radiation-produced point defects by dislocations [29, 30] or other extended defects [33] differently oriented with respect to the external stress. In these models it is assumed that under irradiation Frenkel pairs of vacancies and SIAs are created in the bulk and annihilate at extended defects (EDs) that are thus considered as *sinks* for freely migrating point defects and their small mobile clusters. Consequently, these models can yield a temperature independent irradiation creep only when the bulk recombination of point defects is negligible, which is not the case under sufficiently low irradiation temperature.

An alternative approach has been proposed in ref. [17], which takes into account that EDs can act as *sources* for the production of *radiation-induced* Schottky defects that do not exist in the bulk. A Schottky defect is a single vacancy or SIA (or a small defect cluster), which can be emitted from the ED surface and which does not require a counterpart of opposite sign in contrast to the bulk production of Frenkel pairs, in which the total numbers of vacancies and SIAs *must be equal*. Results of molecular dynamics (MD) simulations [37] have shown that more vacancies than SIAs can be produced in the vicinity of the dislocation cores, and the required energy, $E_v$, is lower than the threshold energy for Frenkel pair production in the bulk.

In this chapter we develop further a mechanism of irradiation creep [17], which is based on the *radiation and stress induced preference in emission* (RSIPE) of vacancies from dislocations differently oriented with respect to the external stress. It is essentially temperature independent in contrast to the SIPA mechanism. Let us make a clear distinction between the creep mechanisms based on the absorption of Frenkel defects and emission of Schottky defects.

### 2.1 SIPA and SIPE mechanisms of creep

For the sake of simplicity, we shall consider the creep rate along the tensile stress axis in a crystal containing two families of straight edge dislocations perpendicular to the external load axis, namely, dislocations with a Burgers vector parallel (type A) or perpendicular (type N) to this axis and presenting densities of $\rho_d^A$ and $\rho_d^N$, respectively. Then the creep rate is simply related to a dislocation climb velocity, $V_d$, which is determined by the difference in absorption of SIAs and vacancies and by emission of vacancies[1] [17]:

$$\dot{\varepsilon} = \rho_d^A b V_d^A = \rho_d^N b V_d^N \tag{1}$$

$$bV_d^Y = D_i \bar{c}_i Z_i^Y - D_v \bar{c}_v Z_v^Y + D_v c_v^{eq,Y} Z_v^Y, \qquad Y = A, N, \tag{2}$$

where $b$ is the magnitude of the Burger's vector, $D_i$ and $D_v$ are the diffusion coefficients of SIAs and vacancies, respectively, $\bar{c}_i$, $\bar{c}_v$ are their mean concentrations, $Z_{i,v}^Y$ is the capture efficiency of dislocations for vacancies (subscript "v") and SIAs (subscript "i"), and $c_v^{eq,Y}$ is the equilibrium concentration of vacancies in a crystal containing Y-type dislocations. The point defect concentrations obey the following rate equations

$$\frac{d\bar{c}_i}{dt} = K_{FP}(1-\varepsilon_i) - k_i^2 D_i \bar{c}_i - \beta_r(D_i + D_v)\bar{c}_i \bar{c}_v, \quad K_{FP} = k_{eff} K, \tag{3}$$

$$\frac{d\bar{c}_v}{dt} = K_{FP}(1-\varepsilon_v) - k_v^2 D_v(\bar{c}_v - \bar{c}_v^{eq}) - \beta_r(D_i + D_v)\bar{c}_i \bar{c}_v, \tag{4}$$

$$\bar{c}_v^{eq} = \frac{Z_v^A \rho_d^A c_v^{eq,A} + Z_v^N \rho_d^N c_v^{eq,N}}{Z_v^A \rho_d^A + Z_v^N \rho_d^N}, \quad k_{i,v}^2 = Z_{i,v}^A \rho_d^A + Z_{i,v}^N \rho_d^N, \tag{5}$$

---

[1] Emission of SIAs from dislocations is not considered due to relatively high activation energies of SIA formation in metals.

where $K$ is the displacement rate, measured in displacements per atom (dpa) per second, $k_{eff}$ is the cascade efficiency for production of stable Frenkel pairs in the bulk, which determines the strength of the source of freely migrating point defects, $K_{FP}$; $\varepsilon_{i,v}$ is the fraction of point defects formed in the in-cascade clusters, $\beta_r$ is a bulk recombination constant and $k_{i,v}^2$ is the dislocation sink strength for SIAs (subscript "$i$") or vacancies (subscript "$v$").

Usually, in the technologically relevant temperature range, steady-state conditions are attained at very early irradiation stages, in which the point defect production is balanced by their loss to sinks, so that $d\bar{c}_{i,v}/dt = 0$, and one obtains from (3) and (4) the relation

$$D_v \bar{c}_v = \frac{k_i^2}{k_v^2} D_i \bar{c}_i + D_v \bar{c}_v^{eq} \tag{6}$$

Thus, the creep rate (1) can be written as

$$\dot{\varepsilon} = \rho_d^A D_i \bar{c}_i \left( Z_i^A - \frac{k_i^2}{k_v^2} Z_v^A \right) + \rho_d^A D_v Z_v^A \left( c_v^{eq,A} - \bar{c}_v^{eq} \right), \tag{7}$$

and, accounting for (2) - (6), one obtains:

$$\dot{\varepsilon} = \frac{\rho_d^A \rho_d^N}{k_v^2} D_i \bar{c}_i \left( Z_i^A Z_v^N - Z_i^N Z_v^A \right) + \frac{\rho_d^A \rho_d^N Z_v^A Z_v^N}{k_v^2} D_v \left( c_v^{eq,A} - c_v^{eq,N} \right) = \dot{\varepsilon}_{SIPA} + \dot{\varepsilon}_{SIDE} \tag{8}$$

The first term in (8) is determined by the stress-induced preference in *absorption* of SIAs and vacancies by dislocations of different types (SIPA mechanism). It is proportional to the difference in absorption (or capture) efficiencies by dislocations of different types, $Z_i^A Z_v^N - Z_i^N Z_v^A$, which is a function of applied stress and material parameters, and to the mean steady-state flux of SIAs, $D_i \bar{c}_i$, that depends on the irradiation dose rate and temperature as follows:

$$D_i \bar{c}_i = -\frac{1}{2} \frac{D_v}{\beta_r} \frac{k_v^2}{k_i^2} \left( k_i^2 + \beta_r \bar{c}_v^{eq} \right) + \left( \frac{1}{4} \left( \frac{D_v}{\beta_r} \frac{k_v^2}{k_i^2} \right)^2 \left( k_i^2 + \beta_r \bar{c}_v^{eq} \right)^2 + K_{FP} \frac{D_v}{\beta_r} \frac{k_v^2}{k_i^2} \right) \tag{9}$$

The product $D_i \bar{c}_i$ is essentially temperature independent and proportional to the dose rate when the bulk recombination is neglected, and it rapidly decreases with decreasing temperature in the recombination dominant region:

$$D_i \bar{c}_i \approx \begin{cases} \dfrac{K_{FP}}{k_i^2}, \dfrac{K_{FP} \beta_r}{D_v k_i^2 k_v^2} \to 0, & \text{sink region} \\[2mm] \left( \dfrac{K_{FP} D_v}{\beta_r} \right)^{1/2}, \dfrac{K_{FP} \beta_r}{D_v k_i^2 k_v^2} \gg 1, & \text{recombination region} \end{cases} \tag{10}$$

The capture efficiencies of dislocations under tensile stress, $\sigma$, are given in the conventional SIPA models [29,30, 36] by the following expressions

$$Z_n^Y = \frac{2\pi}{\ln(R_d/R^Y)}, \qquad R_d = \sqrt{\frac{1}{\pi\rho_d}}, \qquad \rho_d = \rho_d^A + \rho_d^N, \tag{11}$$

$$R_n^Y = \frac{1}{2}L_n\left(1 + A_n^Y \frac{\sigma}{\mu}\right), \qquad L_n = \frac{\mu b(1+\nu)}{3\pi k_B T(1-\nu)}|\Omega_n|, \quad n = i,v, \; Y = A,N \tag{12}$$

where $R_d$ is the radius of a cylindrical "region of influence" around a straight dislocation, which is determined by the dislocation density, $R_n^Y$ is the effective "capture radius" of a dislocation determined by its elastic interaction with the point defects, $\mu$ is the shear modulus of the matrix, $\nu$ is the Poisson ratio, $\Omega$ is the point defect relaxation volume, and $A_n^Y$ is a function of the dislocation orientation, $Y$, and on the differences in shear modulus, $\mu$, and bulk modulus, $\kappa$, between the matrix and point defects. The maximum SIPA effect is obtained in the extreme case of $\mu_i \to 0$, $\kappa_i \to \infty$ for SIAs and $\mu_v \to \mu$, $\kappa_v \to 0$ for vacancies: $A_v^Y = 0$, $A_i^A \approx 0.1$, $A_i^N \approx -0.1$. In this case, the external stress does not influence the dislocation capture efficiency for vacancies ($Z_v^A = Z_v^N \equiv Z_v$), while it increases the SIA capture efficiency of A type dislocations and decreases the SIA capture efficiency of N type dislocations. In the first order of approximation for $\sigma/\mu \ll 1$ one obtains from (8) that the SIPA creep rate is proportional to the applied load:

$$\dot{\varepsilon}_{SIPA} \approx \frac{\rho_d^A \rho_d^N}{k_v^2} D_i \bar{c}_i Z_v \left(A_i^A - A_i^N\right)\frac{\sigma}{\mu} \le 0.2 \frac{\rho_d^A \rho_d^N}{k_v^2} D_i \bar{c}_i Z_v \frac{\sigma}{\mu}. \tag{13}$$

The second term in (8) is determined by the stress-induced preference in *emission* (SIPE) of vacancies by dislocations of different types. It is proportional to the difference between equilibrium vacancy concentrations corresponding to dislocations of different types. Without irradiation, the equilibrium concentrations can be obtained from thermodynamics by minimizing the free energy of a crystal containing identical dislocations [17]:

$$c_v^{th,N} = \exp\left(-\frac{E_v^{f,d}}{k_B T}\right) \equiv c_v^{th}, \quad c_v^{th,A} = \exp\left(-\frac{E_v^{f,d} - \sigma\omega}{k_B T}\right) = c_v^{th} \exp\left(\frac{\sigma\omega}{k_B T}\right) \tag{14}$$

where $E_v^{f,d}$ is the vacancy formation energy at a free dislocation, and $\omega$ the atomic volume. Substituting (14) in (8) one obtains the creep rate due to the *temperature and stress induced preference in emission* of vacancies (TSIPE) (known also as classical Nabarro creep mechanism) that is a special example of a more general SIPE mechanism. In the first order of approximation for $\sigma\omega/k_B T \ll 1$ one obtains that the TSIPE creep rate is proportional to the applied load and to the mean steady-state flux of *thermally-produced* vacancies, $D_v c_v^{th}$, that depends exponentially on temperature:

$$\dot{\varepsilon}_{TSIDE} \approx \frac{\rho_d^A \rho_d^N Z_v^A Z_v^N}{k_v^2} D_v c_v^{th} \frac{\sigma\omega}{k_B T} \tag{15}$$

The temperature dependence of the net creep rate due to both SIPA and TSIPE mechanisms given by eqs. (8) - (15) is shown in Fig. 1 versus experimentally measured irradiation creep compliance (the net part of creep per 1 MPa stress and irradiation dose of 1 dpa, which is independent from the void swelling), which is known to be about $10^{-6} MPa^{-1} dpa^{-1}$ for a large range of austenitic steels irradiated in nuclear reactors in a wide temperature interval [32].

It can be seen that the SIPA mechanism, even at the best choice of material parameters, predicts much lower creep rates than experimentally observed. Besides, the SIPA creep rate is proportional to the mean SIA concentration, which decreases with decreasing temperature (see eq. (10)). On the other hand, the TSIPE mechanism based on the *thermal emission* of vacancies does not depend on irradiation, but it depends exponentially on temperature and becomes inefficient at T < 0.5 Tm (Tm is the melting temperature). However, when we consider how irradiation can enhance the rate of vacancy emission from dislocations and how this enhancement depends on the dislocation type, then we will obtain an alternative mechanism of irradiation creep based on the *radiation and stress induced preference in emission* of vacancies (RSIPE). In the following section we describe possible mechanisms of radiation-induced emission of vacancies from dislocations.

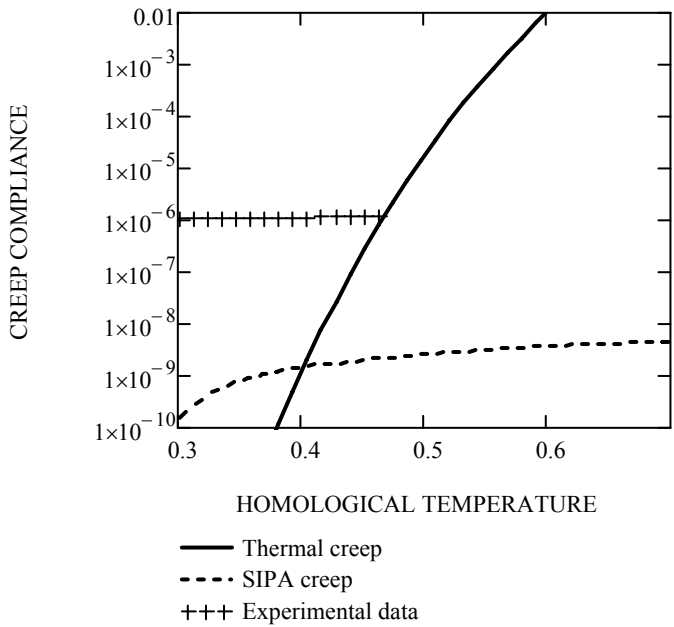

Figure 1. SIPA and TSIPE creep compliance versus experimentally observed irradiation creep compliance [32]. The dislocation density is $\rho_d^A = \rho_d^N = 10^{14} m^{-2}$. Other material parameters are given in Table 1.

### 2.2 Radiation-induced emission of vacancies from extended defects

It is known that not all the energy of the primary knock-on atom (PKA) is spent for the production of stable defects. A considerable part of the PKA energy is spent for production of unstable Frenkel pairs (UFPs) [4, 5] and mobile lattice vibrations, such as focusons [6-9] and quodons [10-14], which can interact with dislocations and other extended defects and produce Schottky defects in their vicinity [20]. Consider the mechanisms of such interaction.

### 2.2.1 Unstable Frenkel pairs

According to MD simulations [37], vacancies can be produced in the vicinity of dislocation cores, and the required energy, $E_v$, is lower than the threshold energy for Frenkel pair production in the bulk. The underlying mechanism can be understood as follows. The dislocation core is surrounded by a region of a radius $R_{cap}$, in which a point defect is unstable since it is captured by the dislocation athermally [38]:

$$R_{cap}^n = \left(\frac{\mu(1+\nu)}{3\pi(1-\nu)E_n^m}|\Omega_n|\right)^{1/2} b, \quad R_{cap}^i \approx 3b, \quad R_{cap}^v \approx b \tag{16}$$

where $E_n^m$ is the migration energy of point defects. The dislocation capture radius for SIAs is larger than the one for vacancies. If a regular atom in the region $R_{cap}^v < r < R_{cap}^i$ gets an energy $E > E_v$ it may move to an interstitial position, where it can be athermally captured by the dislocation, leaving behind it a stable vacancy. The capture time is about $10^{-11} - 10^{-12} s$ so that the process can be described as an effective emission of a vacancy by the dislocation due to its interaction with an unstable Frenkel pair. Since the energy of the system is increased as a result of vacancy formation and corresponding dislocation climb due to the SIA capture, the minimum transferred energy, $E_v$, should exceed the energy of vacancy formation *accounting for the work due to dislocation climb in the stress field*. N-type dislocations do not interact with a tensile stress, and so one has $E_v^N = E_v^f$, while for A-type dislocations one has $E_v^A = E_v^f - \sigma\omega$, as in the case of the conventional Nabarro climb.

Let us estimate the rate of radiation-induced emission of vacancies from dislocations by UFPs created under electron irradiation. The rate of radiation-induced emission of vacancies from a dislocation is given by the product of the UFP production rate per atom, $K_{UFP}(E_v^Y)$, and the number of atomic sites in the region $R_{cap}^v < r < R_{cap}^i$, from where a SIA can be captured by the dislocation, leaving behind it a vacancy. Then, the rate of vacancy emission per unit dislocation length is given by

$$J_v^Y \approx 2\pi R_{cap}^v R_{iv} \frac{K_{UFP}(E_v^Y)}{\omega}, \quad R_{iv} \equiv R_{cap}^i - R_{cap}^v \tag{17}$$

$$K_{UFP}(E_v^Y) = j_e \int_{E_v^Y}^{E_d} \frac{d\sigma}{dE} dE \tag{18}$$

where $j_e$ is the electron flux, and $d\sigma/dE$ is the differential cross section for producing a PKA of energy $E$, which is written in the McKinley - Feshbach approximation as [0]

$$\frac{d\sigma}{dE} = \pi \left(\frac{Z\varepsilon^2}{m_e c^2}\right)^2 \frac{1-\beta^2}{\beta^4} \frac{E_m}{E^2}\left[1-\beta^2 \frac{E}{E_m} + \frac{\pi\beta Z}{137}\left\{\left(\frac{E}{E_m}\right)^{\frac{1}{2}} - \frac{E}{E_m}\right\}\right], \tag{19}$$

$$\beta = \left[1-\left(1+\frac{E_e}{m_e c^2}\right)^{-2}\right]^{1/2}, \quad E_m = \frac{2E_e(E_e + 2m_c c^2)}{Mc^2}, \tag{20}$$

where $E_e$ is the electron energy, $Z$ is the atomic number, $\varepsilon$ and $m_e$ are the electron charge and mass, respectively, $M$ is the target atom mass, $c$ is the light velocity, $E_d$ is the displacement energy and $E_m$ is the maximum energy that can be transferred to a PKA.

The rate of production of stable Frenkel pairs in the bulk is given by

$$K_{FP} = j_e \int_{E_d}^{E_m(E_e)} \frac{d\sigma}{dE} dE \tag{21}$$

Under low-energy sub-threshold irradiation, i.e. $E_m < E_d$, the Frenkel pair production in the bulk is suppressed ($K_{FP} = 0$) but the unstable Frenkel pairs can still be produced until $E_m > E_v$.

Now we will relate the rate of vacancy emission per unit dislocation length (17) to the radiation-induced vacancy *out-flux* density across the surface surrounding the dislocation vacancy capture region:

$$\left(j_v^Y\right)_{out}^{irr} = \frac{J_v^Y}{2\pi R_{cap}^v} = \frac{R_{iv}}{\omega} K_{UFP}(E_v^Y), \tag{22}$$

Then, the boundary condition of general type at the dislocation capture surface can be written as the difference between the influx and out-flux of vacancies:

$$\left(j_v^Y\right)_{r=R_{cap}^v} = \left(j_v^Y\right)_{in} - \left(j_v^Y\right)_{out} = \frac{v_v}{\omega} c(R_{cap}^v) - \left(j_v^Y\right)_{out} = \frac{v_v}{\omega}\left[c(R_{cap}^v) - \frac{\omega\left(j_v^Y\right)_{out}}{v_v}\right], \tag{23}$$

$$\left(j_v^Y\right)_{out} = \left(j_v^Y\right)_{out}^{irr} + \left(j_v^Y\right)_{out}^{th} = \left(j_v^Y\right)_{out}^{irr} + \frac{v_v}{\omega} c_v^{th,Y} = \frac{v_v}{\omega}\left[c_v^{irr,Y} + c_v^{th,Y}\right] \tag{24}$$

where $v_v \approx D_v/b$ is the vacancy transfer velocity from the matrix to the capture region, and $c(R_{cap}^v)$ is the actual vacancy concentration at the dislocation core surface, which should be obtained from the solution of the diffusion problem accounting for boundary conditions (23) at all dislocations in the system. In equilibrium conditions, which can be realized *without irradiation* or under *sub-threshold* irradiation ($K_{FP} = 0$) in a system with *identical* EDs, e.g. dislocations of one type, there would be *no flux* across their capture surfaces. Then $c(R_{cap}^v)$ will be equal to the equilibrium vacancy concentration, $c_v^{eq,Y}$:

$$c_v^{eq,Y} = c_v^{irr,Y} + c_v^{th,Y}, \tag{25}$$

$$c_v^{irr,Y} = \frac{\omega\left(j_v^Y\right)_{out}^{irr}}{v_v} \approx \frac{\omega\left(j_v^Y\right)_{out}^{irr} b}{D_v} = \frac{bR_{iv}}{D_v} K_{UFP}(E_v^Y), \tag{26}$$

*Without irradiation*, one has simply $c(R_{cap}^v) = c_v^{th,Y}$ given by eq. (14). Under *sub-threshold* irradiation, $c(R_{cap}^v)$ is given by the sum $c_v^{eq,Y} = c_v^{irr,Y} + c_v^{th,Y}$, where $c_v^{irr,Y}$ is due to the *radiation-induced emission* of vacancies and is given by eq. (26). In both cases, the mean steady-state concentration in the system coincides with the equilibrium concentration (no gradients and defect fluxes between the EDs).

Finally, under over-threshold irradiation ($K_{FP} > 0$), the flux across the capture surface is proportional to the difference between the mean steady-state and equilibrium concentrations: $(j_v^Y)_{r=R_{cap}^v} \propto \bar{c}_v - c_v^{eq}$.

The calculated vacancy concentrations for electron irradiations of different energies are presented in Fig. 2. It can be seen that the radiation-induced equilibrium concentration dominates completely over the thermal one at temperatures below 0.5 $T_m$. At low temperature end, the thermal equilibrium concentration is practically zero, whereas the radiation-induced equilibrium concentration increases with decreasing temperature due to decreasing diffusion coefficient, and may exceed the melting point level. It saturates at very low temperatures, where the radiation-induced diffusion coefficient, $D_v^{irr} \approx K_D b^2$, dominates over the thermal one, $D_v^{th} = D_v^0 \exp\left(-\frac{E_v^m}{k_B T}\right)$, where $E_v^m$ is the vacancy migration activation energy, and a frequency of radiation-induced "jumps", $K_D$, is given by

$$K_D = j_e \int_{E_v^m}^{E_m(E_e)} \frac{d\sigma}{dE} dE, \qquad (27)$$

The mean steady-state concentration is larger than the equilibrium concentration between 0.3 and 0.6 $T_m$ (swelling region) but both concentrations are exactly the same under sub-threshold irradiation in a system with identical EDs (Fig. 2b: true equilibrium state).

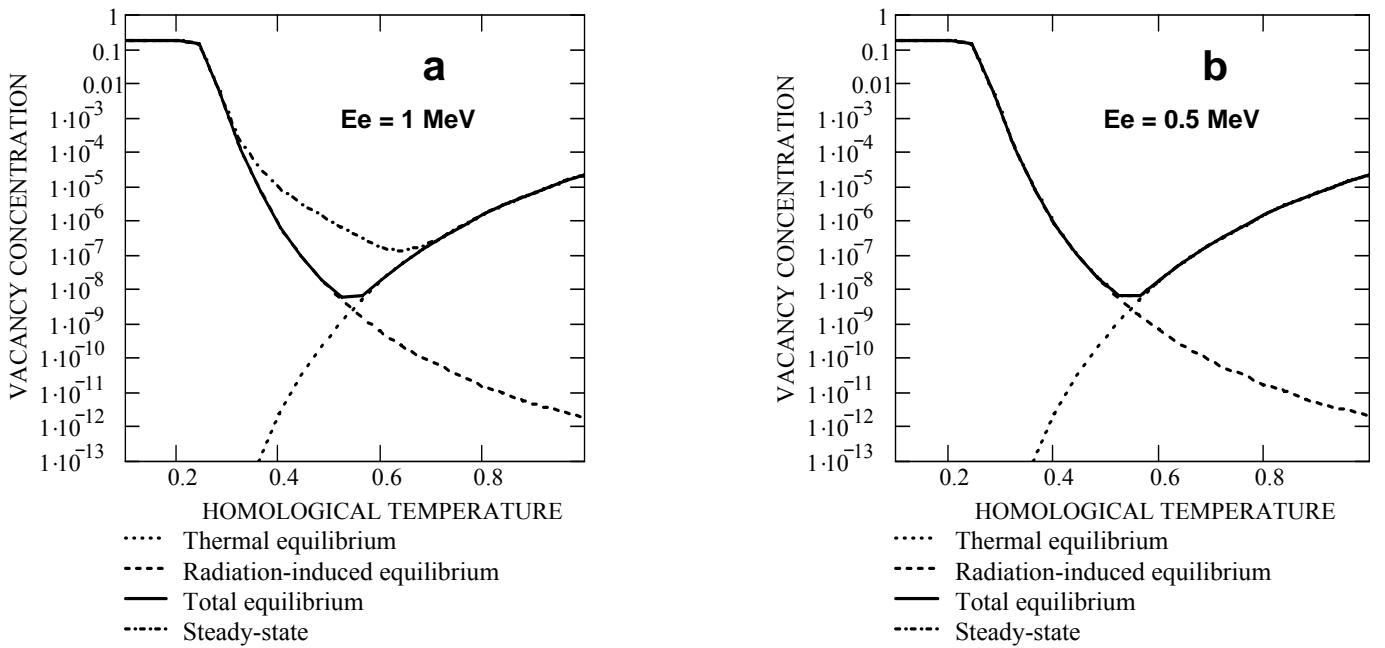

Figure 2. Equilibrium and steady-state vacancy concentrations in Ni (a) under over-threshold ($E_e = 1 MeV$, $j_e = 10^{19} e/cm^2$, which corresponds to $K_{FP} = 2 \times 10^{-4} s^{-1}$) and (b) sub-threshold ($E_e = 0.5 MeV$, $K_{FP} = 0$) electron irradiations. In the second case, the steady-state and equilibrium concentrations coincide. Dislocation density, $\rho_d^A = 10^{14} m^{-2}$. Other material parameters are given in Table 1.

### 2.2.2 Focusons

Focusons [6-9] are produced in the recoil events and *transfer energy* along close packed directions of the lattice, but there is *no interstitial transport* by a focusing collision, which enlarges their range considerably as compared to that of a *crowdion*. The energy range in which focusons can occur has an upper limit $E_F$, which has been estimated to be about 60 eV for Cu, 80 eV for Ag and 300 eV for Au [9]. In an ideal lattice a focuson travels in a close packed direction and loses its energy continuously by small portions, $\varepsilon_F$, in each lattice spacing, which determines its propagation range, $l_0(E, E_v)$, as a function of initial energy, $E$, and the final energy, $E_v^Y$, required for the vacancy ejection [9]:

$$l_0(E, E_v^V) = l_F^0 \ln(E/E_v^V), \quad l_F^0 = b/\varepsilon_F, \quad \varepsilon_F \approx 0.1 \div 0.01 \tag{28}$$

The initial focuson energy is completely transformed into lattice vibrations or heat. However, if a focuson has to cross a region of lattice disorder, a defect may be produced in its surroundings [8]. The presented derivation of the radiation-induced equilibrium concentration can be extended to include the focuson mechanisms of Schottky defect formation. To do so let's estimate the rate of vacancy emission from a cylinder of the radius $R_{cap}^v$ and unit length surrounding a dislocation core due to its interaction with incoming focusons, $J_v^Y$. It is equal to the number of energetic focusons ($E > E_v^Y$) absorbed by the dislocation core from the bulk per unit time, which is proportional to the focuson production rate per atom and the number of atomic sites from which a focuson can reach the dislocation core.

If a dislocation line is oriented along one of the $n_c$ close-packed directions, focusons can reach a dislocation core from $n_c - 2$ cylinders of a radius $R_{cap}^v$ and length $l_0(E, E_v)$ protruded along the closed packed directions. Accordingly, the rate of the vacancy emission due to absorption of focusons is given by

$$J_v^Y = \frac{2R_{cap}^v(n_c - 2)}{\omega n_c} \int_{E_v^Y}^{E_F} l_0(E/E_v^Y) z(E_p, E) dE \tag{29}$$

where $E_F$ is the focuson maximum energy and $z(E_p, E) dE$ is the number of focusons produced by a primary recoil atom (PKA) with energy, $E_p > E_F$, in the energy range $(E, dE)$:

$$z(E_p, E) = \frac{2E_p}{E_F^2} \ln \frac{E_F}{E}, \quad E \leq E_F, \tag{30}$$

The rate of production of focusons can be expressed via the displacement rate per atom, *K*:

$$K_F = K_{PKA} \int_{E_v^Y}^{E_F} z(E_p, E) dE = 4K \frac{E_d}{E_F} \int_1^{E_v^V/E_F} \ln x \, dx \tag{31}$$

where $K_{PKA} = K(2E_d/E_p)$ is the rate of PKA production and $E_d$ is the displacement energy.

In the equilibrium state, the vacancy flux out of the dislocation core (29) is balanced by the vacancy flux into the core, which is given by the product of the local vacancy concentration, $c_v^{irr,Y}$, the cylinder surface, $2\pi R_{cap}^v$, and the rate, at which a vacancy cross the capture surface, $v_v$:

$$J_v^Y = \frac{2\pi}{\omega} R_{cap}^v c_v^{irr,Y} v_v \tag{32}$$

Equalizing eqs. (29) and (32) one obtains the following expression for the focuson-induced equilibrium concentration of vacancies at a dislocation core:

$$D_v c_v^{irr,Y} = b l_F^0 K_F^d(E_v^Y), \quad K_F^d(E_v^Y) \equiv K \frac{4(n_c - 2)}{\pi n_c} \frac{E_d}{E_F} \int_1^{E_v^V/E_F} \ln x \ln\left(x \frac{E_F}{E_v^V}\right) dx \approx \frac{1}{2} K_F \tag{33}$$

which is very similar to the expression (26), but the effective production rate of focusons, $K_F^d(E_v^Y)$, stands here for the UFP production rate, and the focuson propagation range, $l_F^0$ stands for $R_{iv} \approx b$. It can be shown that the focuson mechanism of vacancy production is more efficient than the UFP one if $l_F^0 > 5 R_{iv}$, which is likely to be the case at sufficiently low temperatures. Focusons are unstable against thermal motion because they depend on the alignment of atoms. Typically, at elevated temperatures, the focuson range is limited to several unit cells and their lifetime is measured in picoseconds [9]. However, there exists much more powerful, essentially non-linear, mechanism of the lattice excitations having large lifetimes and propagation distances, which is called *discreet breathers* (DBs) or *quodons* [10-14].

*2.2.3 Quodons*

According to molecular dynamic (MD) simulations, DBs are highly anharmonic vibrations, being sharply localized on just a few sites, which have frequencies above or below the phonon band, and so they practically don't interact with phonons [10]. Breathers can be sessile or mobile depending on the initial conditions. The relative atomic motions in a breather can be of transverse or longitudinal type. Quodons, by definition, are high energy mobile longitudinal optical mode discreet breathers. As the incident focuson energy is dispersed, on-site potentials and long range co-operative interactions between atoms can influence the subsequent dispersal of energy in the lattice by the creation of quodons that don't interact with phonons and thus are thermally stable. Russell and Eilbeck [11] have presented evidence for the existence of energetic, mobile, highly localized quodons that propagate great distances in atomic-chain directions in crystals of muscovite, an insulating solid with a layered crystal structure. Specifically, when a crystal of muscovite was bombarded with alpha-particles at a given point at 300 K, atoms were ejected from remote points on *another* *face of the crystal*, lying in atomic chain directions at more than $10^7$ unit cells distance from the site of bombardment.

Although these results relate to layered crystals there is evidence that quodons can occur in non-layered crystals, but with shorter path lengths of order $10^4$ unit cells. This was reported in connection with radiation damage studies in silicon [12] and with diffusion of interstitial ions in austenitic stainless steel [13]. This points out to the possibility of vacancy emission from any extended defect (void, edge dislocation or grain boundary) in a process of quodon-induced energy deposition.

Let us modify the model with account of quodon-induced production of Schottky defects. Assuming that quodons and focusons can eject vacancies from EDs independently by a similar mechanism, one can write the focuson and quodon -induced equilibrium concentration of vacancies as a sum:

$$D_v c_v^{irr,Y} = bl_F^0 K_F^d \left(E_v^Y\right) + bl_Q^0 K_Q^d \left(E_v^Y\right), \qquad (34)$$

where $K_Q$ and $l_Q^0 >> l_F^0$ are the effective production rate and the propagation range of quodons in a perfect lattice, respectively. Determination of the quodon production rates and ranges requires a detailed model of their formation and propagation in real crystals, which is lacking. So we will use the quodon propagation range as a free parameter and assume below that the quodon production rate is equal to the focuson production rate given by eq. (33). Its dependence on the focuson energy is shown in Fig. 3. It can be seen that in the energy range required for the Schottky defect formation, the rate of focuson production is close to the displacement rate.

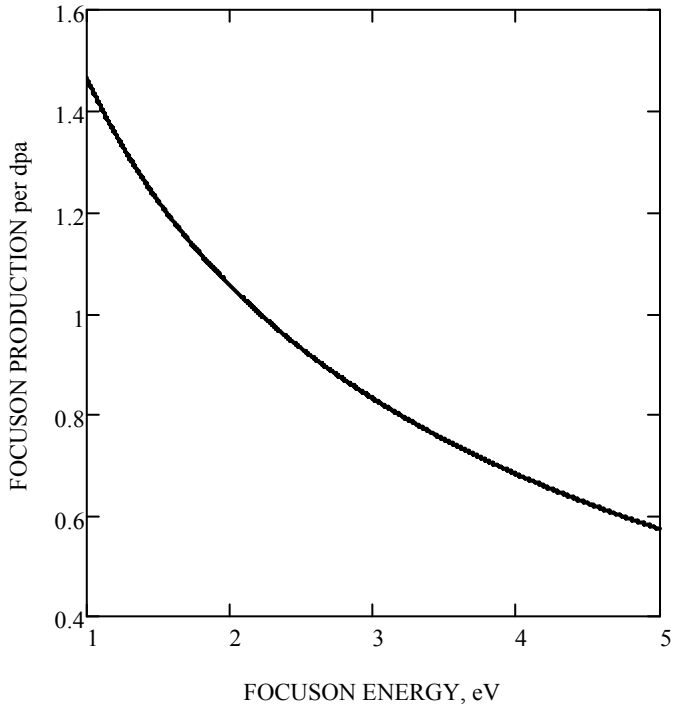

Figure 3. The focuson production rate per dpa as a function of the focuson energy given by eq. (33)

## 2.3 Creep due to radiation and stress induced preference in emission (RSIPE)

In a system with different EDs, such as A-type and N-type dislocations, the true equilibrium state is impossible to reach even without irradiation, due to the difference in equilibrium concentrations at A-type and N-type dislocations, which is a driving force for thermal creep as pointed out in the previous section. Under irradiation, this difference starts to depend on the irradiation dose rate, resulting in a new mechanism of creep, based on *radiation and stress induced preference in emission* of vacancies (RSIPE). Substituting eqs. (26) and (34) into the expression for SIPE creep rate (8) one obtains:

$$\dot{\varepsilon}_{SIPE} = \dot{\varepsilon}_{TSPE} + \dot{\varepsilon}_{RSIPE}, \tag{35}$$

$$\dot{\varepsilon}_{RSIPE} \approx \frac{\rho_d^A \rho_d^N Z_v^A Z_v^N}{k_v^2} D_v c_v^{irr,N} \frac{\sigma\omega}{E_v^{f,d}} \tag{36}$$

$$D_v c_v^{irr,N} = bR_{iv} K_F^d\left(E_v^{f,d}\right) + bl_F^0 K_F^d\left(E_v^{f,d}\right) + bl_Q^0 K_Q^d\left(E_v^{f,d}\right), \tag{37}$$

where the terms in eq. (37) correspond to the UFP, focuson, and quodon mechanisms of vacancy emission, respectively. In this derivation, we have taken into account that each constituent in the radiation-induced vacancy emission is inversely proportional to the vacancy formation energy: $D_v c_v^{irr,N} \propto 1/E_v^{f,d}$, $D_v c_v^{irr,A} \propto 1/(E_v^{f,d} - \sigma\omega)$. Considering the first order of approximation for $\sigma\omega/E_v^f \ll 1$ one has $D_v c_v^{irr,A} \approx D_v c_v^{irr,N}(1 + \sigma\omega/E_v^f)$, which results in a creep rate (36) proportional to the applied load, which is very similar to the expression for thermal creep (15) but $c_v^{irr,N}$ stands here for $c_v^{th,N}$ and $E_v^f$ stands for $k_B T$.

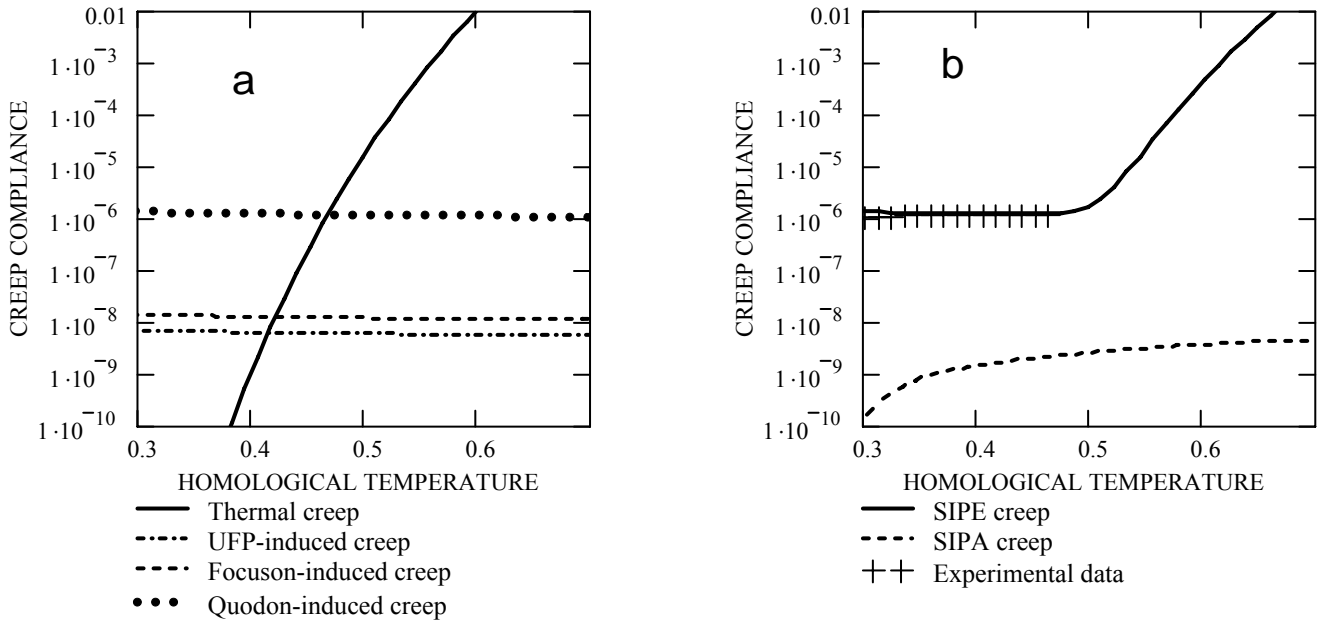

Figure 4. Comparison of different creep mechanisms. (a) Creep rates due to different SIPE mechanisms: thermal; UFP ($R_{iv} = b$); focuson ($l_F^0 = 10b$); quodon ($l_Q^0 = 1000b$). (b) SIPA and SIPE (total) creep compliance versus experimentally observed irradiation creep compliance [0]. The dislocation density is $\rho_d^A = \rho_d^N = 10^{14} m^{-2}$. Other material parameters are given in Table 1.

Comparison of different SIPE mechanisms (Fig. 4a) shows that an efficiency of each one is proportional to the range of the corresponding lattice excitation, which is expected to be much higher for quodons than for focusons or UFPs. The RSIPE creep rate agrees perfectly with experimental data (Fig. 4b) assuming the quodon propagation range to be about $10^3$ unit cells, which is well within the range previously reported for metals [13]. In contrast to SIPA, the RSIPE creep rate does not depend on production and recombination of Frenkel pairs, and so it can be temperature independent as demonstrated in Fig. 4. For over-threshold irradiation it is proportional to the displacement rate:

$$\dot{\varepsilon}_{RSIDE}(E_m > E_d) \approx \rho_d b l_Q^0 \frac{E_d}{E_F} \frac{\sigma \omega}{E_v^{f,d}} K, \quad \rho_d^A = \rho_d^N = \rho_d, \tag{38}$$

which explains why the experimentally measured creep compliance is about $10^{-6} MPa^{-1} dpa^{-1}$ for a large range of austenitic steels irradiated in nuclear reactors in a wide temperature interval [32].

*2.4 SIPE vs. SIPA summary*

It is interesting to note that in the first models of irradiation creep, (see e.g. ref. [36]) it was suggested that below 0.5 $T_m$ the thermal self-diffusion coefficient in the expression for *thermal creep* rate could be replaced by a larger value, namely, by the product of the diffusion coefficient with a steady-state concentration of vacancies increased by the irradiation. This model was criticized because irradiation was thought to enhance the vacancy concentration *in only one way*, i.e. by producing Frenkel pairs in the bulk, which plays no role in *emission* driven mechanisms. We have considered another way of increasing the vacancy concentration by irradiation, which is based on their enhanced emission from extended defects. This radiation-induced emission can be different for dislocations of different orientations in the stress field due to variations in the vacancy formation energy, which are proportional to the external stress. Hence, there are some similarities between RSIPE and TSIPE creep models, which differ qualitatively from SIPA models based on the *long-range* interaction of point defects with dislocations.

The second difference is that the bulk recombination of point defects is efficient in suppressing SIPA creep (Fig. 4b), while the SIPE creep is not affected by the recombination.

The third difference between the RSIPE and SIPA models is their different dependence on the electron beam energy as has been pointed out in [17]. RSIPE creep should be observed under sub-threshold irradiation, when SIPA creep is zero, which could be used for an unambiguous experimental verification of the RSIPE mechanism.

## 3 IRRADIATION SWELLING

A schematic representation of the radiation effect on perfect and real crystals is shown in Fig. 5.

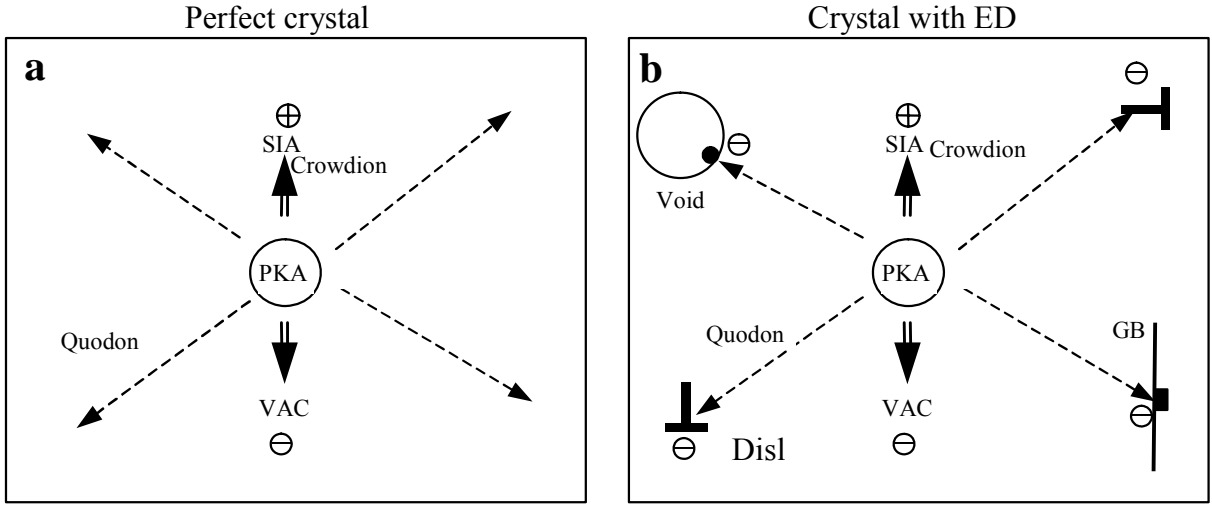

Figure 5. Illustration of Frenkel and Schottky defect production in perfect and real crystals. (a) Frenkel pair formation in the bulk by short-ranged crowdions: SIA is the self-interstitial atom, VAC is the vacancy, arrows show the propagation directions of quodons that do not produce defects in ideal lattice; (b) vacancy formation at extended defects due to interaction of long-ranged quodons with voids, dislocations (⊥) and grain boundaries (GB).

Irradiation produces stable Frenkel pairs by a crowdion mechanism, which is a major driving force of the radiation damage. Even in a perfect crystal (Fig. 5b), a considerable part of energy is spent on production of quodons[2] and only subsequently dissipates to heat. In a real crystal (Fig. 5b), if a quodon has to cross a region of lattice disorder, a Schottky defect may be produced in its vicinity, which is another driving force of microstructural evolution. In the previous section, we have considered one example of its action, i.e. irradiation creep due to the interaction of quodons with dislocations of different orientations in respect to the stress field. A similar effect can be shown to arise due to the interaction of quodons with grain boundaries (GBs) of different orientations in respect to the stress field, which may be important for description of irradiation creep of ultra fine grain materials. In this section, we will consider another prominent example, namely, irradiation swelling due to the void formation, and will show some important consequences of the quodon interaction with voids. We will start with a description of experimental observations of the radiation-induce void "annealing", which requires special irradiation conditions, since under typical irradiation, this phenomenon is obscured by the void growth due to absorption of vacancies produced in the bulk. Our approach is based on the suppression of radiation-induced defect formation in the bulk. This can be achieved if the irradiation temperature is decreased following the void formation. The bulk recombination of Frenkel pairs increases with decreasing temperature suppressing the production of freely migrating vacancies (the driving force of void growth). On the other hand, the rate of radiation-induced vacancy emission from voids remains essentially unchanged, which can result in the void dissolution. Two types of irradiating particles (heavy ions and protons) and irradiated materials (Ni and Cu) will be described in this chapter.

---

[2] Quodons are defined here as quasi-one-dimensional lattice excitations that transfers energy along close-packed directions, which applies also to a classic focuson [4-9].

### *3.1 Experimental observations of the radiation-induced void annealing*

### *3.1.1 Irradiation of nickel with Cr ions*

Nickel foils of 100 micron thickness have been irradiated with 1.2 MeV Cr ions at 873 K up to the total ion fluence of $10^{21}$ m$^{-2}$, which corresponded approximately to the irradiation dose of 25 displacements per atom (dpa) at the dose rate, $K = 7 \times 10^{-3} s^{-1}$ [19]. Examination of control samples in transmission electron microscope (TEM) has revealed formation of a high number density (~$10^{21}$ m$^{-3}$) of voids of 40-50 nm in diameter. The remaining foils have been irradiated subsequently up to the ion fluence of $10^{21}$ m$^{-2}$ at two different temperatures, 798 K and 723 K, respectively. The resulting microstructure is shown in Fig. 6, from which it is evident that the irradiation at lower temperatures has made the voids to decrease in size. The quantitative analysis of the void swelling confirms this conclusion: the void swelling has decreased by a factor of ~ 5 (Fig. 6).

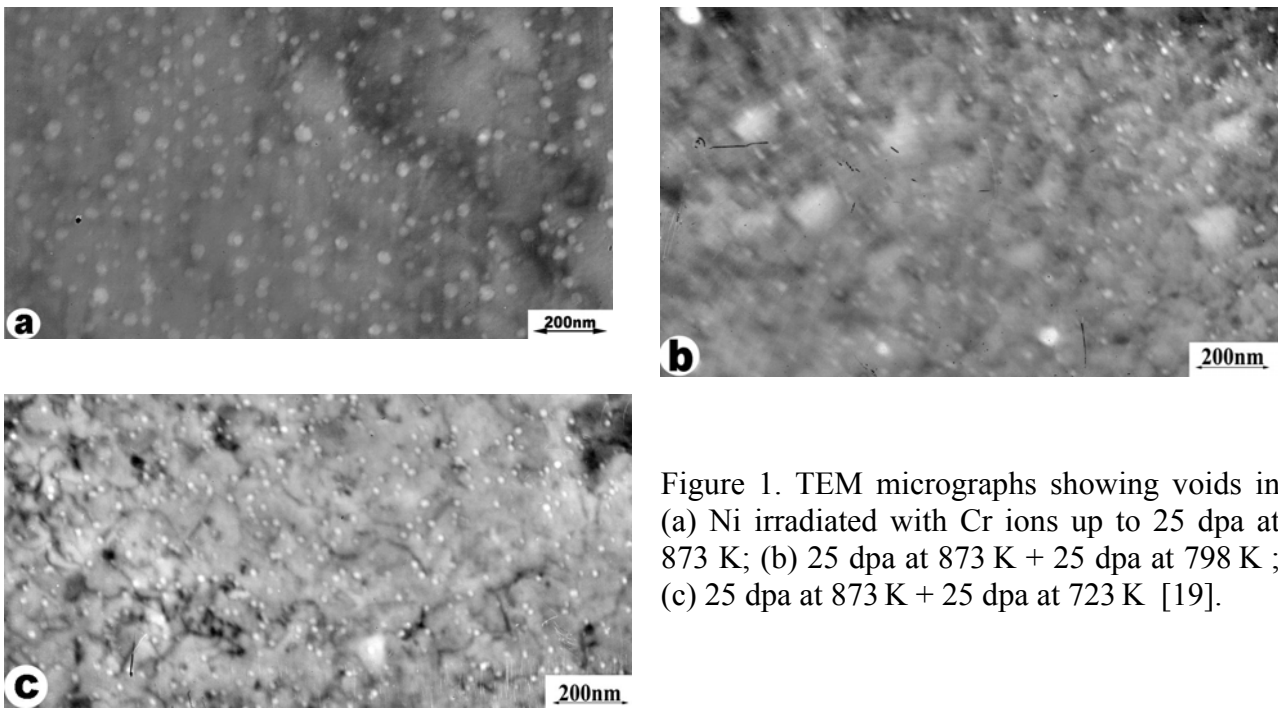

Figure 1. TEM micrographs showing voids in (a) Ni irradiated with Cr ions up to 25 dpa at 873 K; (b) 25 dpa at 873 K + 25 dpa at 798 K ; (c) 25 dpa at 873 K + 25 dpa at 723 K [19].

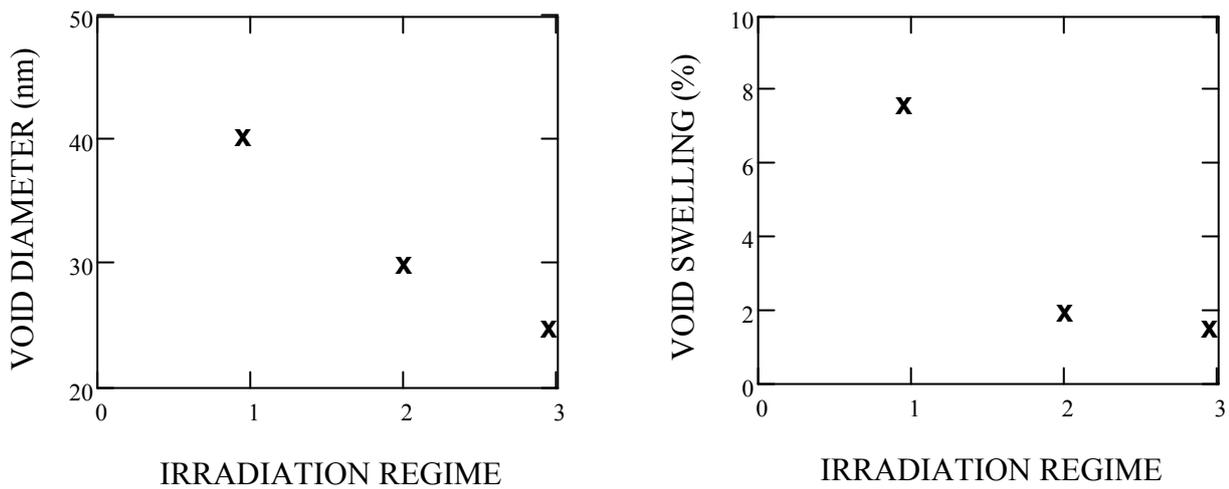

Figure 6. Mean void diameter and swelling in Ni irradiated with Cr ions up to 25 dpa at 873 K (regime 1); 25 dpa at 873 K + 25 dpa at 798 K (regime 2) and 25 dpa at 873 K + 25 dpa at 723 K (regime 3). The size of symbols "x" corresponds to the mean error in void measurements [19].

### *3.1.2 Irradiation of nickel with protons*

In the second type experiment, nickel films of 100 nm thickness have been irradiated with 30 keV protons at 873 K up to the total proton fluence of $10^{22}$ m$^{-2}$, which corresponded to the irradiation dose of 6 dpa at the dose rate of $K = 6 \times 10^{-4} s^{-1}$ and to the concentration of implanted hydrogen of 14 at.%. TEM examination of the control samples has revealed formation of a bimodal distribution of cavities (Fig. 7) consisting of a high number density (~$10^{22}$ m$^{-3}$) of compact hydrogen bubbles of 10-15 nm in diameter and a relatively low density population of vacancy voids, which have irregular shapes and sizes ranging from 20 to 25 nm, which is consistent with the conventional view on the bubble-void transition effects in metals irradiated with gas ions [39]. Some of the films have been irradiated subsequently up to a maximum proton fluence of $2 \times 10^{22}$ m$^{-2}$ at 723 K. Fig. 7b shows that the low temperature irradiation resulted in an increase of the number density of small bubbles and disappearance of large vacancy voids. As a result, the net swelling has decreased by a factor of 3 (Fig. 8).

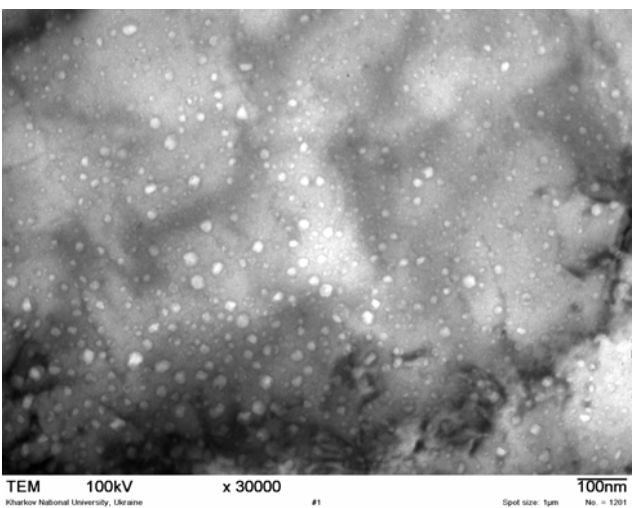
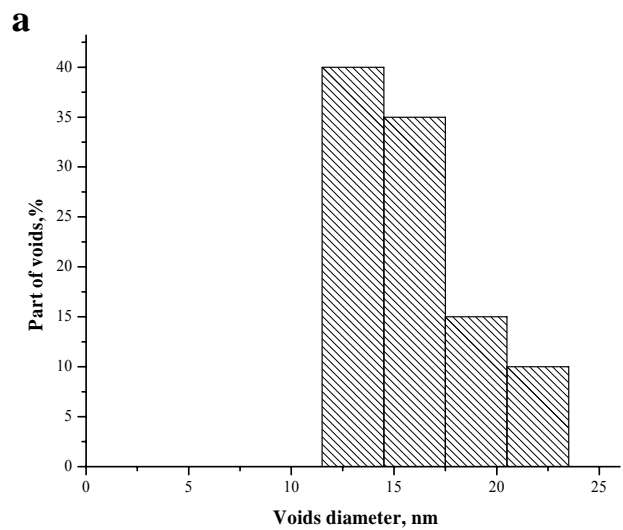

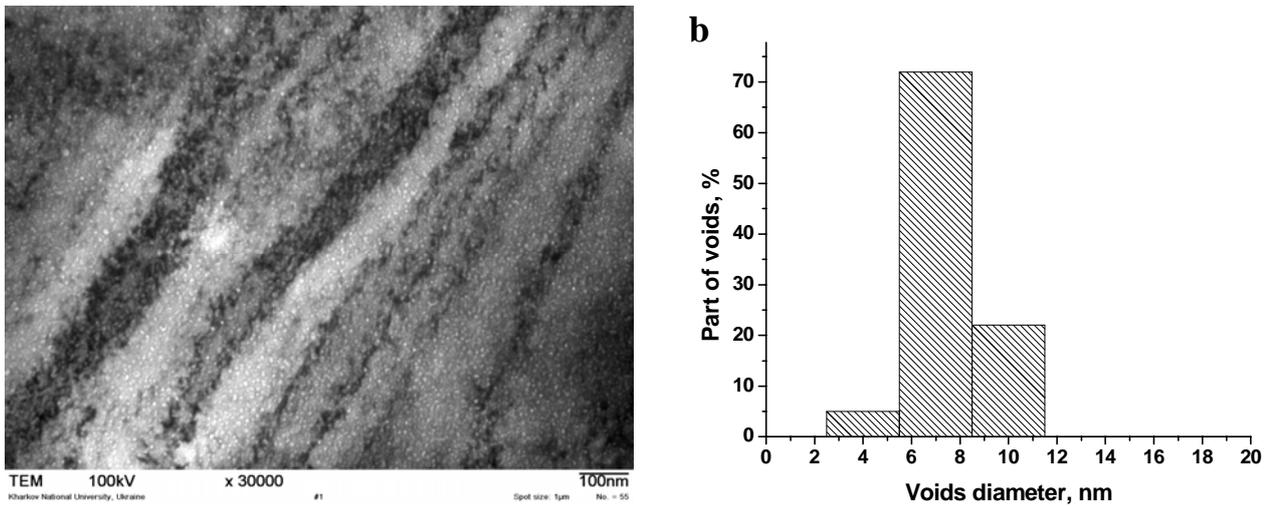

Figure 7. TEM micrographs and corresponding size histograms of cavities in nickel irradiated by 30 keV protons at different temperatures. Picture (a) shows a mixed population of voids and hydrogen bubbles after initial irradiation up to 3 dpa at 873 K. Picture (b) shows a high density population of hydrogen bubbles after subsequent irradiation up to 6 dpa at 723 K.

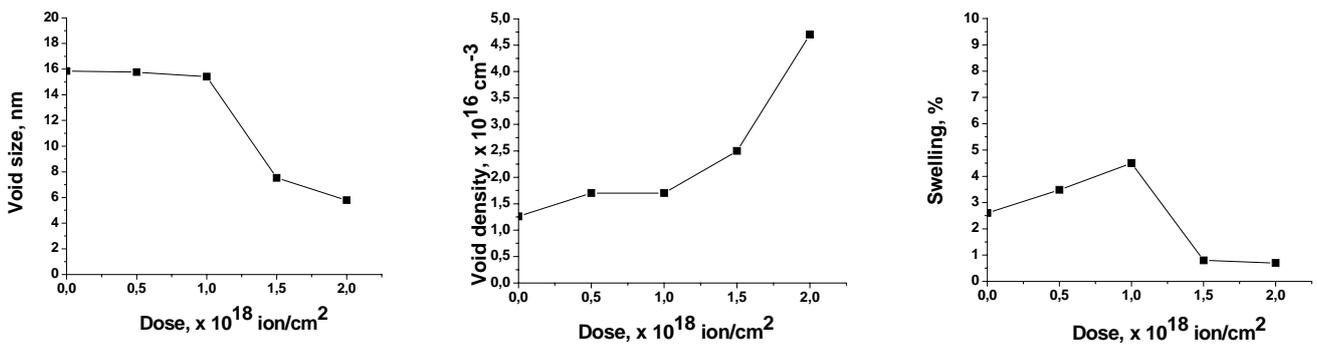

Figure 8. Void size, number density and swelling in nickel pre-irradiated by 30 keV $H^+$ ions at 873 K as a function of irradiation dose at 723 K.

*3.1.3 Irradiation of copper with protons*

In the third type experiment, copper films of 100 nm thickness have been irradiated with 30 keV protons at 773 K up to the total proton fluence of $10^{22}$ m$^{-2}$, which corresponded to the irradiation dose of 10 dpa at the displacement rate, $K = 10^{-3} s^{-1}$, and to the concentration of implanted hydrogen of 14 at.%. TEM examination of the control samples has revealed formation of a bimodal distribution of cavities (Fig. 9) consisting of a high number density (~$10^{21}$ m$^{-3}$) of hydrogen bubbles of 20-40 nm in diameter and a relatively low density population of vacancy voids in the size range of 60-80 nm. Some of the films have been irradiated subsequently up to a maximum proton fluence of 6x$10^{21}$ m$^{-2}$ at 573 K. Fig. 9b shows that the low temperature irradiation resulted in a gradual increase of the number density of small bubbles and shrinkage of large vacancy voids. As a result, the net swelling has decreased by a factor of 3 (Fig. 10).

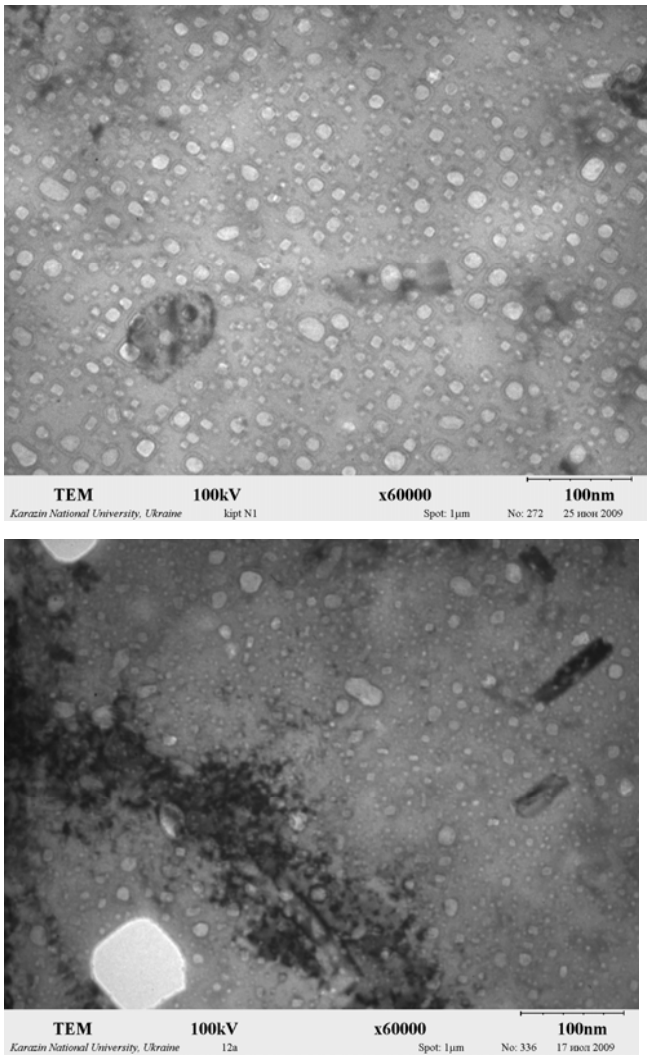
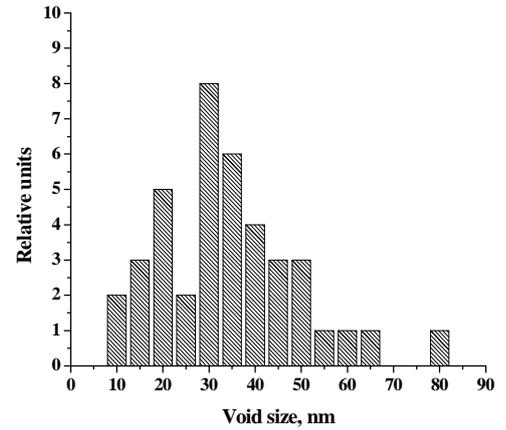
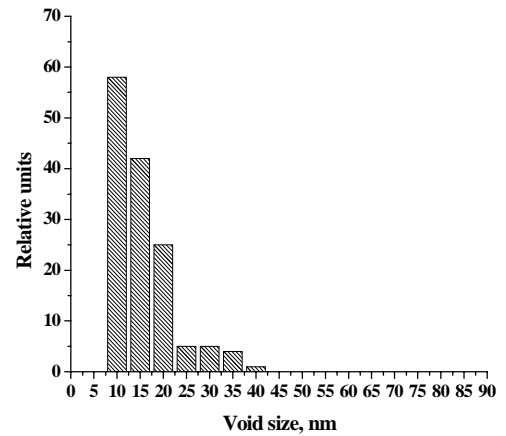

Figure 9. TEM micrographs and corresponding size histograms of cavities in copper irradiated by 30 keV protons at different temperatures. Picture (a) shows a mixed population of voids and hydrogen bubbles after initial irradiation up to 10 dpa at 773 K. Picture (b) shows the microstructure after subsequent irradiation up to 6 dpa at 573 K.

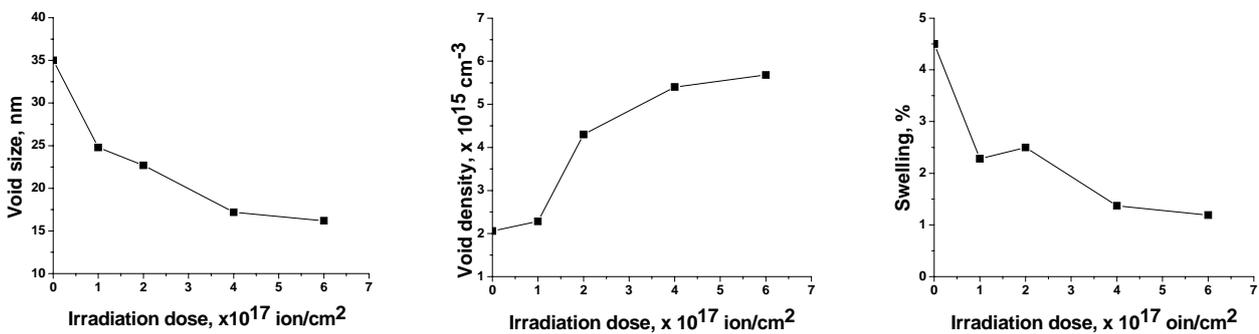

Figure 10. Void size, number density and swelling in copper pre-irradiated by 30 keV $H^+$ ions at 773 K as a function of irradiation dose at 573 K.

A few previous experiments have indicated that voids can shrink with decreasing temperature after they have been formed under more favorable conditions [41, 42]. According to Steel and Potter [42], voids formed during 180 keV $Ni^+$ ion bombardment of Ni at 923 K shrink rapidly when subjected to further bombardment at temperatures between 298 and 823 K. The authors have attempted to explain the observations using the rate theory modified to include the additional metal interstitial atoms injected by the ion beam. However, this effect was shown to be negligible due to a very low "production bias" introduced by injected SIAs (about 0.1%) [19]. What is more, in the experiment [19] described above, the energy of Cr ions was an order of magnitude higher than that in Ref. [42]. Accordingly, the Cr ions came to rest at a distance of about $10^4$ nm, which exceeded the depth, at which voids have been produced, by orders of magnitude. In the case of the proton irradiation of Ni and Cu, there were no additional metal interstitial atoms, but a high concentration of hydrogen bubbles was formed, which were stable under low temperature irradiation in contrast to voids. The underlying physical mechanisms for the observed radiation-induced annealing of voids are considered in the following section.

### 3.2 *Quodon model of the radiation-induced void annealing*

The energy of vacancy formation at the void surface is lower than that at a flat stress-free surface because of the surface tension, which „tries" to make a void shrink. On the other hand, the vacancy formation energy at the bubble surface is higher than that at the flat stress-free surface because of the gas pressure, which opposes the surface tension [40]. The same is true for the vacancy formation at the dislocation core due to the stalking fault energy. It means that the energy delivered by quodons to the void and bubble surfaces and dislocation cores ejects more vacancies from voids than from bubbles or dislocations. The fate of a void or a bubble depends on the balance between the absorption and emission of vacancies, which may become negative for voids when the absorption of vacancies formed in the bulk decreases. This is what happens at lower irradiation temperature due to enhancement of the bulk recombination of vacancies with SIAs.

In order to incorporate this qualitative picture in the modified rate theory, let us evaluate the *local equilibrium concentration of vacancies* at the surface of a cavity (void or gas bubble), $c_v^{eq,C}$, which is determined by the rates of vacancy emission due to thermal or radiation-induced fluctuations of energy states of atoms surrounding the void. Generally, $c_v^{eq,C}$ is given by the sum of the thermal and the radiation-induced constituents:

$$c_v^{eq,C} = c_v^{irr,C} + c_v^{th,C}, \qquad c_v^{th,V} = \exp\left(-\frac{E_v^f + P_g \omega - 2\gamma\omega/R}{k_B T}\right) \qquad (39)$$

where $E_v^f$ is the thermal vacancy formation energy at a free surface, $\gamma$ is the surface energy, $P_g$ is the gas pressure inside a gas bubble. In order to evaluate the radiation-induced constituent, $c_v^{irr,C}$, let's estimate the number of vacancies ejected from a void of the radius $R$ due to its interaction with incoming quodons, $dN_v/dt$. Similar to the dislocation case (29), it is equal to the number of energetic focusons ($E > E_v^C$) absorbed by the cavity from the bulk per unit time, which is proportional to the quodon production rate per atom and the number of atomic sites from which a quodon can reach the cavity:

$$J_v^C = \frac{4\pi R}{\omega} \int_{E_v^V}^{E_F} l_0(E/E_v^C) z(E_p, E) dE \qquad (40)$$

In the equilibrium state the vacancy flux out of the cavity (40) is balanced with the vacancy flux into the cavity, which is given by the product of the local vacancy concentration, $c_v^{irr,C}$, the cavity surface, $4\pi R$, and the rate, at which a vacancy cross the cavity surface, $v_v$:

$$J_v^C = \frac{4\pi}{\omega} R c_v^{irr,C} v_v \qquad (41)$$

Equalizing eqs. (40 and 41) one obtains, similar to eq. (33), the following expression for the quodon-induced equilibrium concentration of vacancies at a cavity surface:

$$D_v c_v^{irr,C} = b l_Q^0 K_Q^C(E_v^C), \quad K_Q^C(E_v^C) \equiv K \frac{E_d}{E_F} \int_1^{E_v^C/E_F} \ln x \ln\left(x \frac{E_F}{E_v^C}\right) dx \qquad (42)$$

where $l_Q^0$, is the quodon propagation range in ideal crystal, $K_Q^C(E_v^C)$, is the effective production rate of quodons. The quodon-induced equilibrium concentrations of vacancies at voids, gas bubbles and dislocations are different due to the difference in the vacancy formation energies (see Table 1), which a physical origin of the driving "force" that tries to minimize the total energy of the system and recover it from the damage being produced by Frenkel defects.

Now the cavity growth (or shrinkage) rate is given by the usual expression [18, 40]

$$\frac{dR_C}{dt} = \frac{1}{R_C}\left(D_v \bar{c}_v Z_v^C - D_i \bar{c}_i Z_i^C - D_v c_v^{eq,C} Z_v^C\right), \qquad C = V, B, \qquad (43)$$

where $Z_{i,v}^C$ is the cavity capture efficiency for vacancies [40] (superscript "V" stands for voids and "B" for bubbles), $\bar{c}_{i,v}$ are the mean concentrations of point defects determined by the rate equations (3) and (4), in which $k_{i,v}^2$ is the sink strength of all microstructure components that determine mean concentrations of Frenkel and Schottky defects:

$$\bar{c}_v^{eq} = \frac{Z_v^d \rho_d^d c_v^{eq,d} + Z_v^C N_C \bar{R}_C \bar{c}_v^{eq,C}}{k_v^2}, \qquad k_{i,v}^2 = Z_v^d \rho_d^d + Z_v^C N_C \bar{R}_C, \qquad (44)$$

where $N_C$, $\bar{R}_C$ are the cavity concentration and the mean radius, respectively. The product $D_v c_v^{eq,C} Z_v^C$ in eq. (43) determines the rate of the void shrinkage due to emission of thermal and radiation-induced Schottky defects. The latter does not depend on temperature in contrast to the difference $D_v \bar{c}_v Z_v^C - D_i \bar{c}_i Z_i^C$, which is due to the biased absorption of Frenkel defects that decreases with decreasing irradiation temperature due to enhancement of the point defect recombination in the bulk. As a result, the net growth rate may become negative below some temperature (or above some dose rate [18]).

Fig.11 shows the temperature dependence of the void growth/shrinkage rate in Ni and Cu irradiated under the present irradiation conditions calculated for quodon and focuson ranges assuming the quodon production rate to be equal to the focuson production rate given by eq. (42). Other material parameters are presented in Table1. The focuson range is limited to about ten unit cells, which would result in negligible deviance from the conventional theory shown as the "classical" limit. The observed void shrinkage with decreasing irradiation temperature can be explained by the quodon model assuming the quodon range to be about $10^3$ unit cells and the quodon-induced vacancy formation energy at dislocations to be 2.215 eV and 1.562 eV in Ni and Cu, respectively. This difference may be due to different stacking fault energies (see Table 1), which makes it more difficult to eject vacancies from dislocations in Ni than in Cu.

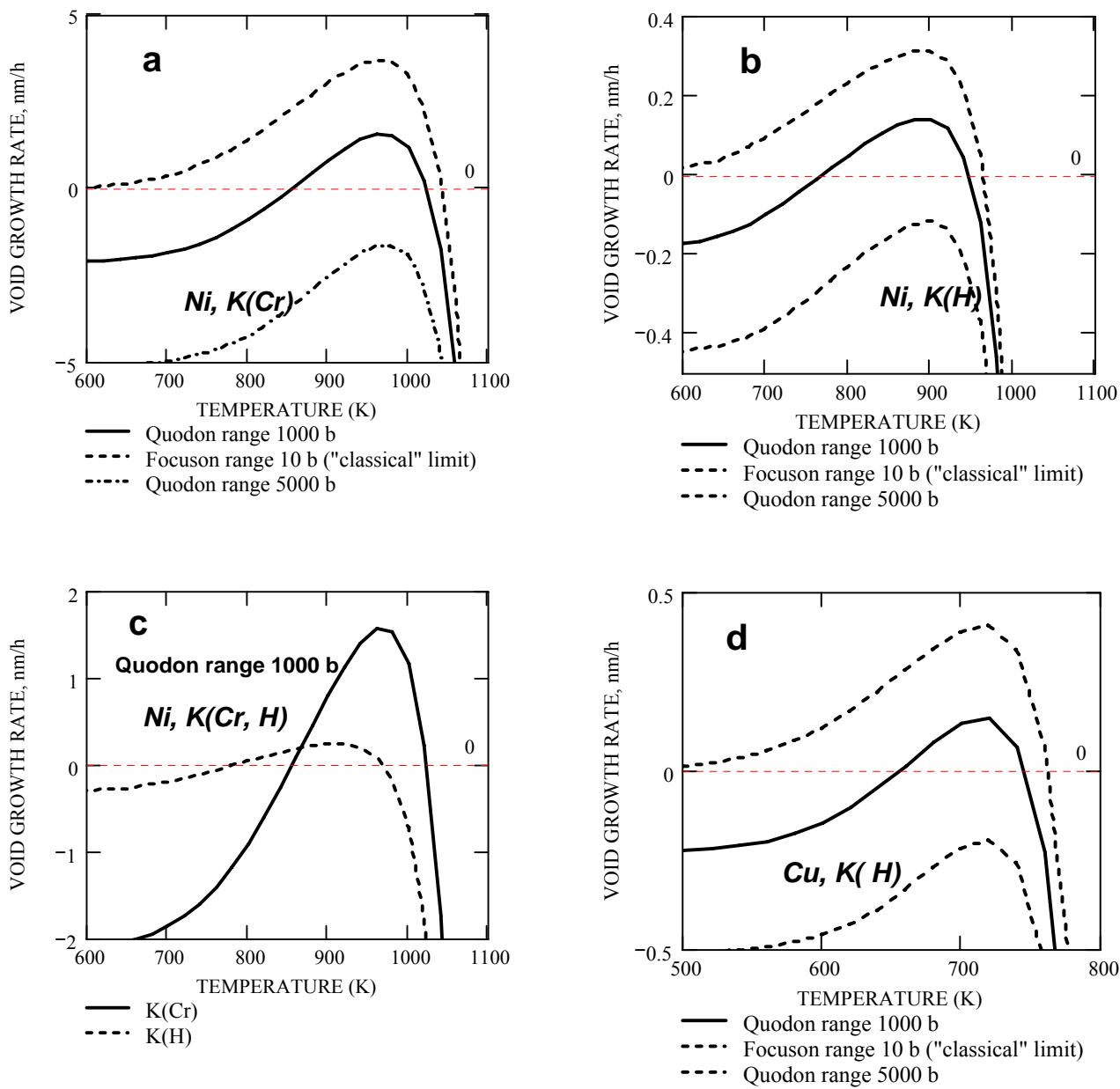

Figure 11. Temperature dependence of the void growth/shrinkage rate in Ni (a), (b), (c) and Cu (d) calculated for different displacement rates: $K(Cr) = 7 \times 10^{-3} s^{-1}$ (a), $K(H) = 6 \times 10^{-4} s^{-1}$ (b), $K(H) = 10^{-3} s^{-1}$ (c). The focuson range is limited to about ten unit cells, which would result in negligible deviance from the conventional theory shown as the "classical" limit.

## 4  VOID LATTICE FORMATION

The radiation-induced void dissolution phenomenon demonstrated in the previous sections is intrinsically connected with a void ordering in a super-lattice, which copies the host lattice [22-27]. The void ordering is often accompanied by a saturation of the void swelling, which makes an understanding of the underlying mechanisms to be both of scientific significance and of practical importance for nuclear engineering. Currently the most popular void ordering concept is based on the mechanisms of anisotropic interstitial transport along the close packed planes [23] or directions [24-27]. In these models, anisotropic diffusion of SIAs [23, 24] or propagation of small SIA loops along the close packed directions [25 -27] makes the competition between growing voids to be dependent on spatial void arrangements. A common difficulty inside the concept of anisotropic SIA transport is the explanation of saturation and even reduction in the void swelling accompanying the void lattice formation [41].

As was suggested in ref. [28], the ordering phenomenon may be considered as a consequence of the energy transfer along the close packed directions provided by long propagating DBs or quodons. If the quodon range is larger than the void spacing, the voids shield each other from the quodon fluxes along the close packed directions, which provide a driving force for the void ordering, as illustrated in Fig. 12.

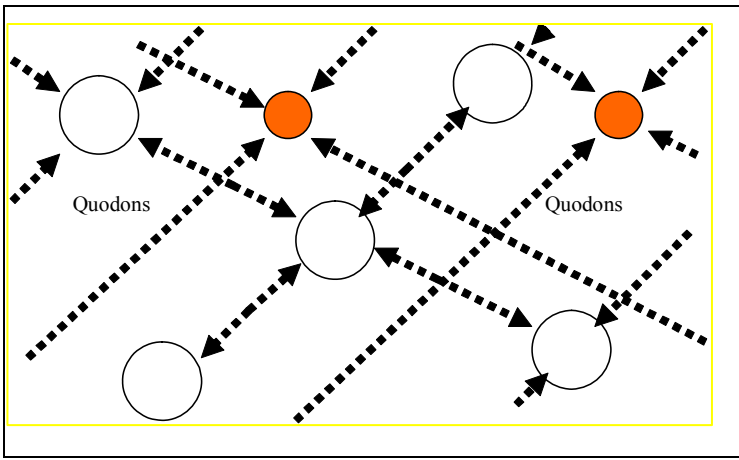

Figure 12. Illustration of the radiation annealing of voids in the "interstitial" positions due to the absorption of quodons coming from larger distances as compared to locally ordered voids that shield each other from the "quodon wind" along the close packed directions.

The emission rate for "locally ordered" voids (which have more immediate neighbors along the close packed directions) is smaller than that for the "interstitial" ones, and so they have some advantage in growth. In terms of the rate theory, the void solubility limit starts to depend on the void position. For the "locally ordered" voids it is proportional to the mean distance from the nearest neighbors along the close packed directions, $l_V$, while for the "interstitial" voids it is proportional to the quodon path length, $l_Q > l_V$ or to the mean distance from neighbors from next coordination spheres, $\langle l_V \rangle > l_V$. Accordingly, the growth rate of the "interstitial" voids is lower than that of the "ordered" ones and so they become smaller. In this way, quodons point out the "interstitial" voids, while their eventual shrinkage is driven by the mechanism of the radiation-induced competition (coarsening) of the voids of different sizes [43], which makes the smaller ("interstitial") voids shrink away, as has been demonstrated in details by Dubinko et al [25-27]. The mechanism of the radiation-induced coarsening [43] is similar to the classical Ostwald ripening of the voids but it is based on the void bias in absorption of SIAs rather than on the evaporation of vacancies. Due to this mechanism the maximum void number density, $N_V$, under irradiation is limited by the value, $N_V^{\max}$, proportional to the dislocation density and the dislocation to void bias factor ratio. At the "nucleation stage" ($N_V \ll N_V^{\max}$) the void growth rate due to the dislocation bias exceeds the void shrinkage rate due to the vacancy emission, and all voids grow. But eventually, at the "ripening stage" ($N_V \to N_V^{\max}$), the net vacancy gain rate is reduced down to the vacancy loss rate (for the "ordered" voids) or below that (for the "interstitial" ones). Consequently, the "interstitial" voids shrink away while the "ordered" ones form a stable void lattice, in which voids neither grow nor shrink.

Let us estimate the void lattice parameters given by the present mechanism. As the void sink strength increases with time, the rate of the quodon-induced vacancy emission from voids, $(d\bar{c}_v/dt)_Q$ equals the rate of the net vacancy absorption by voids due to the dislocation bias, $(d\bar{c}_v/dt)_B$, thus providing the condition for swelling saturation:

$$(d\bar{c}_v/dt)_Q \approx k_V^2 D_v c_v^{irr}, \qquad (45)$$

where $k_V^2 = 4\pi N_V \bar{R}_V$ is the void sink strength. The bias-induced constituent to the void growth is given by the usual expression (see e.g. [25-27]):

$$(d\bar{c}_v/dt)_B \approx \Delta^* k_V^2 k_D^2 \delta_D / (k_V^2 + k_D^2), \quad \Delta^* \equiv D_v(\bar{c}_v - \bar{c}_v^{eq}) \approx K_{FP}/(k_V^2 + k_D^2), \qquad (46)$$

where $k_D^2$, $\delta_D$ are the dislocation sink strength and the bias factor, respectively, and $\Delta^*$ is the vacancy supersaturation, which is determined by the ratio of the production rate of freely migrating Frenkel pairs, $K_{FP}$, to the microstructure sink strength if the latter dominates over the bulk recombination of Frenkel pairs (see eq. (10)). In this case, equalizing the right sides of eqs. (45) and (46) with account of (42) one obtains a relation between the void and dislocation sink strengths corresponding to the saturation of swelling:

$$(k_V^2 + k_D^2) \approx (K_{FP} k_D^2 \delta_D / K_Q l_Q^0 b)^{1/2}, \qquad (47)$$

The second relation is determined by the radiation-induced coarsening mechanism, which becomes efficient at $N_V = N_V^{\max}$ making the smaller ("interstitial") voids shrink away [25-27]:

$$N_V = N_V^{\max} = \Delta^* \delta_D k_D^2 / 4\pi \alpha_{iv}^*, \qquad (48)$$

$$\alpha_{iv}^* \equiv \alpha_{iv} + D^*\alpha_\gamma/\Delta^*, \quad D^* \cong D_v c_v^{th}, \tag{49}$$

where $\alpha_{iv}^*$ is the coarsening parameter, $\alpha_{iv} \equiv \alpha^{im} + \alpha^d(2\gamma/\mu b)$ is the constant of the void bias due to elastic interaction with point defects, $\alpha^{im}$ and $\alpha^d$ are the constants of image and elasto-diffusion interaction, $\gamma$ is the surface energy, μ is the shear modulus.

Substituting (48) into (47) one obtains both the stable void radius and the number density expressions, from which the ratio of the void lattice parameter (VLP) to the void radius is estimated as follows:

$$a_{VL}/\overline{R} \approx (\alpha_{iv}^*)^{-2/3} (k_D^2 \delta_D)^{1/6} (K_Q l_Q^0 b / K_{FP})^{1/2} \tag{50}$$

Fig. 13a shows the dependence of the VLP/R ratio on the DB path length given by eq. (50) assuming that $k_D^2 \delta_D = 10^{14} m^{-2}$, and the coarsening parameter, $\alpha_{iv}^* \approx \alpha_{iv} \approx 5b$, is determined by material constants, which is true for sufficiently low irradiation temperatures or high dose rates corresponding to the void lattice formation. The tendency of the VLP/R ratio decrease with increasing temperature may be expected from eq. (50) due to the increase of the coarsening parameter, $\alpha_{iv}^*$, with increasing temperature according to eq. (49).

Experimentally observed values of the VLP/R ratio range from about 4 to 15 and have a tendency of decreasing with increasing temperature [25-27]. It can be seen from Fig. 13a that the present estimate of the quodon range is well within the range suggested in ref. [13] and it agrees with the quodon range of $10^3$ unit cells assumed in this chapter to describe experimental observations of the radiation-induced void annealing and irradiation creep.

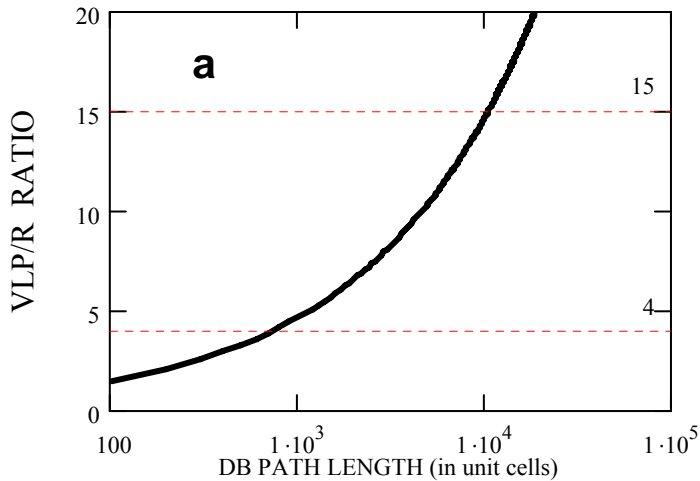

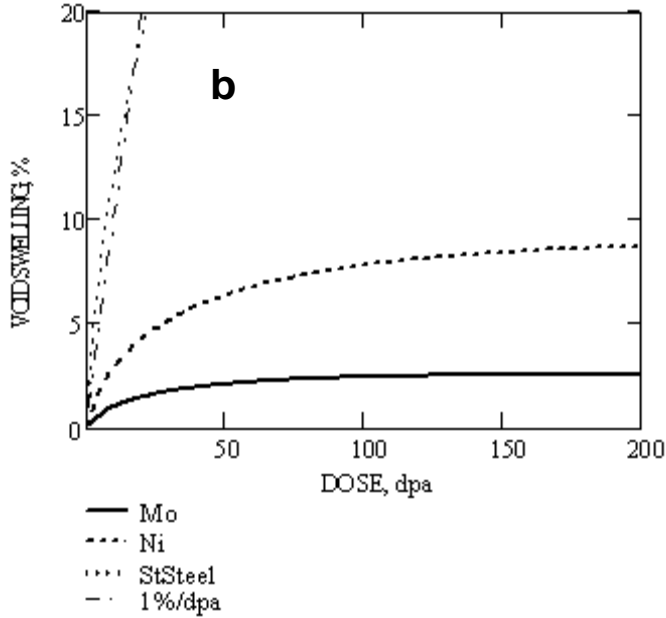

Figure 13. (a) Dependence of the VLP/R ratio on the quodon range according to eq. (50). The markers show the experimentally observed range of VLP/R ratio values in different materials. (b) Dose dependence of swelling in different metals calculated for typical reactor irradiation conditions (displacement rate, $K = 10^{-6} s^{-1}$, T = 673 K) assuming different void number densities, $N_V$ deduced from experimental data: Mo ($N_V = 2 \times 10^{22} m^{-3}$), Ni ($N_V = 8 \times 10^{21} m^{-3}$), St steel ($N_V = 2 \times 10^{21} m^{-3}$).

Finally, the swelling saturation accompanying the void ordering can be achieved due to the competition between the absorption of Frenkel defects and emission of Schottky defects provided that the void concentration is high enough, as it is illustrated in Fig. 13b for metals with different void concentrations.

## 5  CONCLUSION

In this chapter, rate theory modified to take account of quodon-induced reactions has been applied to describe irradiation creep, radiation-induced annealing of voids and the void ordering. The underlying mechanisms seem to have a common nature based on the vacancy emission from voids and dislocations due to their interaction with irradiation-induced quodons. The quodon propagation distance in cubic metals deduced from the comparison between the theory and experiment quodon range is about $10^3$ unit cells, which is consistent with data on the radiation-induced diffusion in stainless steel [13].

In order to forecast the behavior of nuclear materials in real radiation environment one has to know the generation rate of quodons and their propagation range in different crystal structures as the functions of impurity atom concentration and type. The latter factor seems to be of a primary technological importance since it offers a new insight on design of radiation-resistant materials.

**Acknowledgements**

This study has been supported by the STCU grant # 4962.

Table 1. Material parameters used in calculations

| Parameter | Value |
| --- | --- |
| Atomic spacing, $b$, m | $3.23 \times 10^{-10}$ |
| Atomic volume of the host lattice, $\omega$, m$^{-3}$ | $2.36 \times 10^{-29}$ |
| Matrix shear modulus, $\mu$, GPa | 35 |
| Interstitial dilatation volume, $\Omega_i$, | $1.2\omega$, $0.6\omega$, |
| Vacancy dilatation volume, $\Omega_v$, | $-0.6\omega$ |
| Bulk recombination rate constant, $\beta_r$, m$^{-2}$ | $8 \times 10^{20}$ |
| Displacement energy, $E_d$, eV | 30 |
| Cascade efficiency for the stable defect production, $k_{eff}$ | 0.1 |
| Fraction of point defects in the in-cascade clusters, $\varepsilon_{i,v}$ | 0.9 |
| Maximum focuson energy, $E_F$, eV | 60 |
| Vacancy formation energy at a free surface, $E_v^f$, eV | 1.8 (Ni), 1.3 (Cu) |
| Vacancy formation energy at a void surface, $E_v^V$, eV | $E_v^f - 2\gamma\omega/R$ |
| Surface energy, $\gamma$, J/m$^2$ | 2 |
| Stacking fault energy, J/m$^2$ | 0.13 (Ni), 0.03 (Cu) |
| Irradiation-induced vacancy formation energy at dislocations, $E_v^d$ | 2.215 (Ni), 1.562 (Cu) |
| Migration energy of vacancies, $E_v^m$, eV | 1.1 (Ni), 0.98 (Cu) |
| Pre-exponent factor, $D_v^0$ | $10^{-5}$ |